\documentclass[12pt]{iopart}
\usepackage{graphicx}
\usepackage{dcolumn}
\usepackage{bm}

\begin{document}

\title{Structural models for the Si(553)-Au atomic chain reconstruction}

\author{S. Riikonen$^{1,2}$, D. S\'anchez-Portal$^{2,3}$}
\address{
$^1$ Departamento de F\'{\i}sica de Materiales, Facultad
de Qu\'{\i}mica, Universidad del Pa\'{\i}s Vasco, Apdo. 1072, 20080
Donostia-San Sebasti\'an, Spain \\
$^2$Donostia International Physics Centre (DIPC), 
Paseo Manuel de Lardizabal 4,
20018 Donostia-San Sebasti\'an, Spain\\
$^3$Centro Mixto CSIC-UPV/EHU, ``Unidad de F\'{\i}sica
de Materiales", Apdo. 1072, 20080 Donostia-San Sebasti\'an, Spain}
\eads{\mailto{swbriris@sc.ehu.es}, \mailto{sqbsapod@sc.ehu.es}}

\begin{abstract}
Recent photoemission experiments
on the Si(553)-Au reconstruction show a one-dimensional 
band with a peculiar $\sim$${1\over4}$ filling. 
This band could provide an opportunity  
for observing large spin-charge separation if electron-electron 
interactions could be increased. To this end, it is necessary 
to understand in detail the origin of this surface band. 
A first step is the determination of the structure of 
the reconstruction.
We present here a study of several structural models using 
first-principles density functional calculations.
Our models are based on a plausible analogy 
with the similar and better known Si(557)-Au surface, and
compared against the sole structure proposed to date 
for the Si(553)-Au system
[Crain JN {\it et al.}, 2004 Phys. Rev. B {\bf 69} 125401 ].
Results for the energetics and the band structures are given.  
Lines for the future investigation are also sketched.
\end{abstract}

\submitto{Nanotechnology, TNT04 special issue}

\maketitle

\section{Introduction}
When restricted to move in one dimension, electrons have been
predicted to behave in a fundamentally different way than
in three dimensions. The nature of the
low-energy electronic excitations changes dramatically.
The single-particle excitations do not provide a good
description and have to be
replaced by separate spin and charge collective
excitations~\cite{Giamarchi04}. This collective
behaviour leads to several exotic phases at low temperatures. 
However, although these theoretical 
predictions are clear and well-founded, the observation 
of the spin-charge separation and other effects has proven quite elusive. 
The main difficulty is the fabrication of clean and well-ordered arrays 
of one-dimensional systems with electronic properties that can be tailored
and are stable against structural distortions.

Monatomic chains of gold atoms are spontaneously formed 
on several vicinal Si(111) surfaces after the deposition of
gold in the sub-monolayer regime.
Each terrace contains a chain of gold atoms running parallel
to the step-edge. Since the average terrace width can be controlled
by the miscut angle, vicinal substrates provide
an unique route to create regular arrays of monatomic metallic 
wires with tuneable inter-chain distances 
and interactions~\cite{himpsel_review}.
Therefore, these surfaces can be
an ideal playground to 
prove the physics of one dimension and, consequently, have attracted much 
attention in recent years. Some examples 
are the Si(557)-Au~\cite{luttinger4} 
and the Si(111)-(5$\times$2)-Au~\cite{5x2_phase}
reconstructions.

The Angle Resolved Photoemission (ARPES) spectrum of
the Si(557)-Au surface 
is dominated by an almost half-filled one-dimensional 
band that splits in two peaks near the Fermi energy (E$_F$).
This was first assigned to the spin-change separation in 
a one-dimensional metal~\cite{luttinger4}. However, later photoemission 
experiments~\cite{557_bands} and calculations~\cite{daniel1}
seemed to discard this interpretation. In fact, 
according to a recent study combining 
ARPES and Scanning Tunnelling Microscopy (STM),
the surface suffers 
a metal-to-insulator transition at low temperatures~\cite{557_wires}. 
We have
proposed an explanation for these observations
in a recent theoretical paper~\cite{sanchez_riikonen}.
The opening of a gap at E$_F$ is associated with 
several structural distortions.
The metal-insulator
transition, along with the temperature dependence of the STM images, 
relates to the dynamical fluctuations of the silicon atoms
in the step-edge. The observed band splitting is caused by 
the spin-orbit interaction~\cite{sanchez_riikonen}. This 
opens the possibility to 
create or transport spin-polarized currents in these systems, which
might be useful for
future spintronic devices.

The flat Si(111)-(5$\times$2)-Au surface has been shown to exhibit
interesting phenomena as well. The STM images show very prominent
chains formed by silicon adatoms 
with variable 
concentration~\cite{adatom_wires,adatoms,5x2_corr,doping}.
Individual adatoms 
can be manipulated with the STM and
could be used as bits of an atomic-scale memory~\cite{memory}.
The ARPES spectrum of this surface 
shows several bands with a strong one-dimensional 
character close to E$_F$~\cite{1d_states,5x2_bands}. However, the most recent
photoemission data suggest that it
is also non-metallic~\cite{arpes_latest}.

The existence of 
a metallic one-dimensional band with 
a peculiar  $\sim$${1\over4}$ filling has been recently
reported for the Si(553)-Au
surface~\cite{himpsel_review,553}.
This quarter-filled band could create an opportunity
for observing large spin-charge separation if electron-electron
interactions (U) could be increased, for example, using a 
somewhat different
substrate. This is in contrast with half-filled
bands, which are unstable against a Mott-Hubbard transition for
large values of U, preventing the 
observation of a Luttinger metal~\cite{himpsel_review}. 

Unfortunately, the atomic structure
of the Si(553)-Au reconstruction 
is not completely established. A reliable structure is   
necessary for a deep understanding  of its electronic structure.
We present here a study of several structural models using
first-principles density functional calculations.
We propose five models based on our earlier experience with 
the closely related Si(557)-Au 
surface~\cite{daniel1,sanchez_riikonen,daniel2}.
At least within our computational scheme, the new models 
are energetically comparable or even more favourable
than the sole structure 
proposed to date for the Si(553)-Au surface~\cite{himpsel_review}.
Band structures
are analysed and the possible origin of the dispersive
one-dimensional bands in this reconstruction is discussed.
We also give ideas and guidelines for future
research in finding the geometry of these stepped gold-induced silicon 
reconstructions.

\section{Studied structural models}

We have based our investigation of the Si(553)-Au 
reconstruction on a plausible analogy with the closely related
and  better known
Si(557)-Au surface. Both surfaces contain steps along
the $[1\bar{1}0]$ direction. In fact, the
orientation of the chains along the $[1\bar{1}0]$ 
directions seems to be common to all the silicon reconstructions
forming monatomic gold chains~\cite{himpsel_review}.
The Si(557)-Au reconstruction is characterized 
by the following elements (see, for example, 
figure~1 in Ref.~\cite{daniel2}):
{\it i}) Gold atoms occupy substitutional positions in the top
Si layer in the middle of the terraces; 
{\it ii}) the atoms close to the step-edge are 
strongly rebonded, forming a structure reminiscent
of the so-called honeycomb chain (HC)~\cite{honeycomb};
{\it iii}) the structure doubles its periodicity along
the direction of the steps due to a buckling of the atoms
in the step-edge;
{\it iv}) there is also a row of adatoms on each terrace.
This structure has been determined using X-ray diffraction~\cite{557_xray}
and confirmed with {\it ab initio} 
calculations~\cite{himpsel_review,daniel1,daniel2}.
The miscut direction is opposite for the Si(553)-Au 
and the Si(557)-Au surfaces. 
This would open the possibility 
of a different rearrangement of the step-edge in both cases. 
However, we do not 
consider this possibility here, and 
our models for the Si(553)-Au are 
fabricated using the building blocks listed above.

\begin{figure}[h!]
\begin{center}
\includegraphics[width=8.0cm,keepaspectratio]{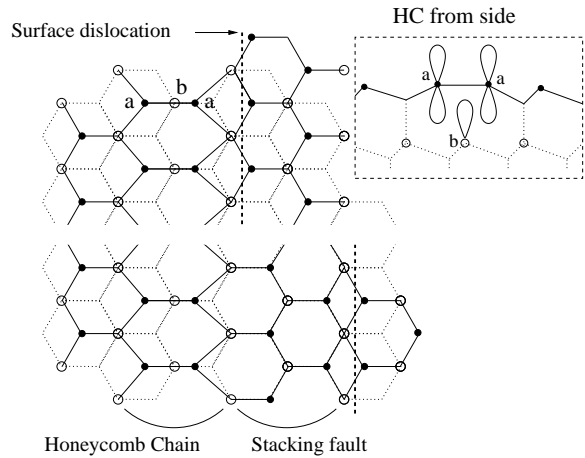}
\end{center}
\caption{Some structural patterns appearing 
in the metal induced reconstructions of the Si(111) surface.
The bonds within the surface bilayer are indicated by 
solid lines while dotted lines are used for the underlying bilayer.
The atoms occupying the highest positions 
in the surface bilayer are represented by solid circles.
Open circles are used for those in the lower bilayer. 
A side view of the HC structure is shown in the inset.
\label{fig:fig1}}
\end{figure}

The importance of the HC structure for the gold induced reconstructions
on vicinal Si(111) has been recently emphasized by 
Crain {\it et al.}~\cite{himpsel_review}. The HC is known to occur
in the Si(557)-Au surface~\cite{daniel1,daniel2,557_xray}. It is also 
the key
ingredient of a recent proposal for the structure of the 
Si(111)-(5$\times$2)-Au reconstruction~\cite{doping}. 
The HC was initially proposed
to explain the low coverage 
reconstructions induced by some alkali metals (Li, Na, K), Mg and Ag
on Si(111)~\cite{honeycomb}.
The HC structure is illustrated in 
figure~\ref{fig:fig1}. We see that the HC structure
involves two unit cells of the unreconstructed Si(111) surface, 
with one atom removed from the top Si layer.
This
flattens the surface and removes surface stress.  
The inset of figure~\ref{fig:fig1} 
shows a side view of the HC.  The unsaturated bonds are 
illustrated,
clarifying the mechanism behind this reconstruction: the bonds
of atoms {\it a} hybridise to create an extra $\pi$-bond.  
This bond is further
hybridised with the dangling-bond of atom {\it b} below. 
The electron count 
of these bonds is odd, so one additional 
electron is needed to yield an electronic closed-shell structure.
If there is a neighbouring alkali-metal atom, it donates one electron 
to the HC, thus contributing to the stabilization 
of the structure~\cite{honeycomb}.
In the case of gold, which has a stronger electron affinity, the situation 
is different.
Gold is likely to take electrons away from the silicon structure.
In principle, this does not prevent an 
electronic stabilization mechanism:
one electron may be transferred to the 6$s$ Au state, leading 
again to a closed-shell structure. However, in reality we are far
from this {\it ionic} situation. The states 
of gold are strongly hybridised with those of 
the neighbouring silicon atoms creating
several dispersive bands that are, 
in principle, 
metallic~\cite{daniel1,sanchez_riikonen}.

The structural elements presented above
can be used to build many models for the gold induced reconstructions
on stepped silicon surfaces.
The length of the terraces and the registry with the underlying 
silicon layers (assumed to be unreconstructed) impose
some geometrical restrictions.
The effect of the introduction of the HC in the Si(111)
is schematically illustrated in 
figure~\ref{fig:fig1}.  While
flattening the surface, the HC creates a stacking fault 
towards the $[\bar{1}\bar{1}2]$ direction ( at the right of
the HC in the figure)
that can be avoided
by introducing a surface dislocation (SD). 
In the upper part of figure~\ref{fig:fig1} 
the SD is introduced immediately 
after the HC, thus recovering the perfect stacking of silicon. 
On the lower part of the figure the stacking 
fault remains until the SD
is introduced.

To name our structural models of the Si(553)-Au surface
we have developed a notation 
which we believe may be useful for other
similar systems.
The first double-layer in the unreconstructed
Si(111) can be expressed using a
1$\times$1 unit cell with a two atoms basis. 
This entity is named ``S''.  If this
block contains a stacking fault (i.e. its bonds 
coincide with those of the second double-layer)
it is called ``S$_f$'', $f$ standing for faulted.  
HC has a 2$\times$1 unit cell
and is indicated by ``H''.  The SD is indicated with ``D''.  
It is basically
equal to ``S'', but one of its atoms is over coordinated.  
Using this notation, the upper part of figure~\ref{fig:fig1}
can be abbreviated as 
(starting from the HC) HDS and the lower part as HS$_f$D.
The substitution of a silicon 
atom by gold is indicated using a $+$ superscript.

Our models for the Si(553)-Au surface can be seen in 
figure~\ref{fig:fig2}. Models I to V are generated by placing 
the HC structure with the accompanying dislocation, and 
the chain of gold atoms in different
positions of the terrace. Model VI has been previously proposed
by Crain {\it et al.}~\cite{himpsel_review}.
The unit cell of the structure VI is three times larger
along the step direction than that of the other models. 
This is due to the
extra Si atoms attached to the step-edge: there are two atoms
every three possible sites. This was proposed as an explanation
for the 1$\times$3 superlattices frequently 
observed in 
the STM images~\cite{himpsel_review,553}. We have 
decided not to impose the appearance of superlattices  
in our models. Instead we studied the stability of the step-edge 
against structural distortions that might also 
be responsible of the STM images like in the case of the
Si(557)-Au surface~\cite{sanchez_riikonen}.

\section{Computational method}

All the calculations were performed using 
the SIESTA code~\cite{siesta1,siesta2}. 
We used the local density approximation~\cite{lda} and
norm conserving pseudopotentials~\cite{tm}, 
which included scalar relativistic
effects for gold~\cite{au1}. 
A double-$\zeta$ (DZ)
basis set of atomic orbitals (i.e. including two 
different 3$s$ orbitals, and another two to represent
the 3$p$ shell) was used for silicon.
The gold basis included doubled and polarized
6$s$ orbitals (thus including a single 
6$p$ shell) and a single 5$d$ shell. For the 
relaxations of some of the most stable structures 
(and the calculation of 
the band structures presented below) we used a double-$\zeta$
polarized (DZP) basis for silicon. 
The energy shift~\cite{siesta2} was chosen to be 200~meV, 
yielding cut-off radii of
5.25 and 6.43 Bohr for the 3$s$ and 3$p$ Si 
orbitals, and 6.24 and 4.51 Bohr for the 6$s$/6$p$ and 5$d$ Au
orbitals, respectively.
                                                                                
We modelled the surface using a finite slab
with the bottom silicon layer saturated with hydrogen.
To avoid artificial stresses the 
lateral lattice parameter was fixed to the bulk theoretical
value calculated with similar approximations. 
The structures were relaxed until the maximum force
component was less than 0.04 eV/\AA.
The distance between neighbouring slabs was 15~\AA. The  
slabs contained either four (systems 
containing one or two Au atoms in the
unit cell) or three (systems with a tripled 
unit cell) silicon bilayers.  
A 4$\times$4 sampling of the surface Brillouin 
zone and a real-space grid equivalent to a 
100~Ry plane-wave cut-off was used.
This guarantees, for a given basis set, 
a convergence of the total energy within $\sim$0.5~meV/\AA$^2$.

\section{Results}

\begin{figure*}
\begin{center}
\includegraphics[keepaspectratio,width=14cm]{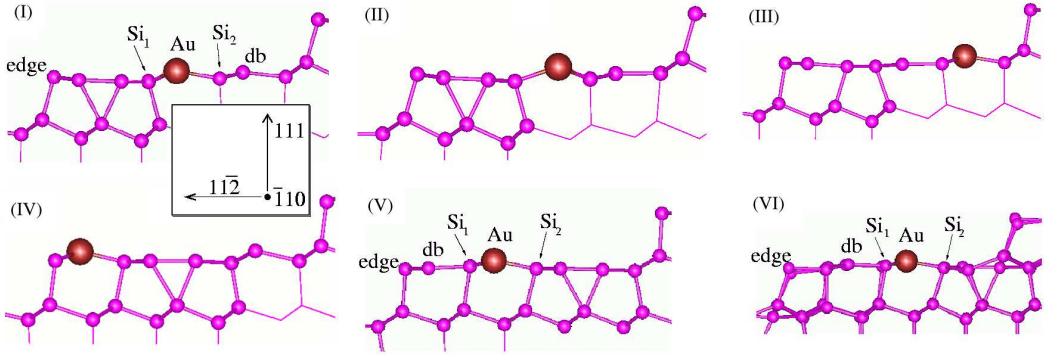}
\end{center}
\caption{Relaxed structures of the different models for 
the Si(553)-Au surface studied here. 
Models I to V are proposed from an analogy with the structure 
of the Si(557)-Au surface. 
Structure VI has been 
proposed recently by Crain {\it et. al.}~\cite{himpsel_review}.
Large circles indicate gold atoms. 
Some of the atoms are labelled (see figure~\ref{fig:fig3}).
The inset clarifies the orientation of the steps.
\label{fig:fig2}}
\end{figure*}

\begin{table}
\caption{Relative surface energies (in meV/\AA$^2$) 
of different structures of the Si(553)-Au surface.
The numbers in the first column
refer to figure~\ref{fig:fig2}, while the names of the models
are assigned according to the nomenclature explained in the 
text.
Column labelled $\times$1 presents the results for the smaller 
unit cell (with a single Au atom).
The systems in columns $\times$2 and $\times$3 included, 
respectively, two and three unit cells
along the step direction. This was done in order to study the stability 
of the step-edge respect to several structural distortions. However, 
the relaxed structures always returned to an almost perfect undistorted
$\times$1 periodicity.
The $\times$1 and $\times$2 slabs contained
four silicon double-layers, while the $\times$3 slabs only
had three silicon double-layers.
The number in parenthesis was obtained using a DZP basis.  
\label{tab:table}}
\begin{indented}
\item[]\begin{tabular}{@{}lllll}
\br
  & 	Model         		& $\times$1	
& $\times$2	& $\times$3 			\\
\mr
I &	HD$^+$S			& 0		
& 0		& 0			\\	%
II &	HS$^+$$_{f}$D		& 1.4		
& -		& -			\\	%
III &	HDS$^+$			& 7.0		
& 7.3		& -			\\	%
IV &	S$^+$HD			& 11.3 		
& -		& -			\\	%
V &	SS$^+$H			& 5.3 (5.6)	
& 6.0		& 5.8			\\	%
\hline
VI & 	Crain {\it et. al.}	& -		
& -		& 5.7	\\
\br
\end{tabular}
\end{indented}
\end{table}
The main results of our work can be found in figure~\ref{fig:fig2}
and table~\ref{tab:table}, where the relaxed structures and
the relative surface energies of the different
models of the Si(553)-Au reconstruction are presented.
Model I is the most stable structure. It presents a strong similarity with 
the Si(557)-Au reconstruction~\cite{daniel2,557_xray}. The HC is placed
at the step-edge and the gold atoms reside in the middle of the terrace.
The main differences in the present case are the presence of a SD,
associated with the
HC and the different orientation of the steps, and 
the absence of adatoms due to the smaller terrace.
Models I and II only differ in the position of the SD
and have almost degenerate energies.
Changes in the location of the gold chain have stronger impact
in the energy. This becomes clear by comparing models I and III, which 
solely differ in the site occupied by the gold atoms. This 
suggests that the ideal location of the gold chain is in the middle of the
terrace, as was previously found 
for the Si(557)-Au reconstruction~\cite{daniel1,daniel2}.

We also tried structures where 
the HC has been moved away from the step-edge.
In both cases, model IV and V, this yields to 
an increase of the surface energy. This seems to indicate
that the HC is a especially stable rearrangement of the step-edge
in vicinal Si(111) surfaces. In fact, in the case
of Si(557)-Au surface, the HC structure formed spontaneously 
at the step-edge during the geometrical optimisations 
using density-functional calculation~\cite{daniel1}. 
The model IV, with the gold chain in a 
position neighbouring to the step-edge, is the most unstable configuration. 
This points
again to the larger stability of the gold substitution in the
middle of the terraces.

The model by Crain \etal~\cite{himpsel_review} (model VI) is 
indeed a
simple variant of model V. A silicon dimer is attached
to two of the step-edge atoms and another two silicon 
atoms in the terrace. 
This creates a $\times$3 modulation 
along the step-edge direction similar to 
what has been observed in some 
STM images~\cite{himpsel_review,553}. Interestingly, the
addition of these silicon dimers leaves the 
surface energy almost unchanged (see table~\ref{tab:table}). 
This is consistent with the observation of large
variations in the concentration of the extra silicon atoms attached 
to the
step edge and, probably, a high mobility of these atoms. Again 
this seems to be consistent with the STM observations.

In the case of the Si(557)-Au surface the doubling of the periodicity 
observed with the STM~\cite{557_bands,557_wires} can be 
explained as a result of a buckling of the 
step edge~\cite{sanchez_riikonen,daniel2}. 
Since this could also provide an explanation for the modulations observed
in the Si(553)-Au surface, 
we have explored the
stability of two of our models (I and V) 
respect to similar distortions.
We have used supercells containing two or three
unit cells along the step. Our relaxations started
from structures where 
one of the atoms of the step edge was
moved either upwards or downwards by
$\sim$0.3~\AA. 
In some cases, the atoms with unsaturated dangling-bonds
in the terrace (see figure~\ref{fig:fig2})
were also moved upwards in an attempt to force their charging 
with electrons transferred from the step edge. In the case
of model I this would leave a metallic step-edge band, susceptible
to favour structural distortions.
In other cases, the whole structure was relaxed under the
constriction of a particular step-edge configuration.
However, in all cases we were unable to stabilize the step-edge buckling.
Once all the degrees of freedom were optimised 
the structures always returned to an almost perfect $\times$1 periodicity. 

We now turn our attention 
to the electronic band structures of our most stable models.
From our studies of the Si(557)-Au surface we 
expect to find two surface bands 
with different dispersions near the Fermi level 
associated with the Si-Au bonds~\cite{daniel1,daniel2}.
A flat band originating from the Si$_2$-Au 
bonds, and a dispersive band coming from the more overlapping 
Si$_1$-Au bonds 
(see figure~\ref{fig:fig2} for the labelling of the atoms).
In the present calculations we have only included scalar-relativistic
effects for gold. In fact, the effect of the spin-orbit interaction 
is expected to be negligible
for the total energy of these surfaces (we have zero spin polarization).
However, we have recently discovered that
the 
dispersive Si$_1$-Au band shows
a considerable spin-orbit splitting (up to $\sim$200~meV
near the Fermi level)~\cite{sanchez_riikonen}.
This is instrumental to reconcile the theoretical band structure
with the photoemission spectra
for the Si(557)-Au surface~\cite{sanchez_riikonen}. 
The photoemission of the 
Si(553)-Au surface shows a half-occupied band featuring 
a clear splitting, almost identical to that observed 
for the Si(557)-Au, and a dispersive 
quarter-filled band at lower binding energies~\cite{553}.
We think that the surface bands showing splitting 
in the experiment have to be identified with those 
coming from the Si$_1$-Au bonds.

\begin{figure*}
\flushleft
\begin{minipage}{7.6cm}
\centering
\includegraphics[keepaspectratio,width=5.0cm]{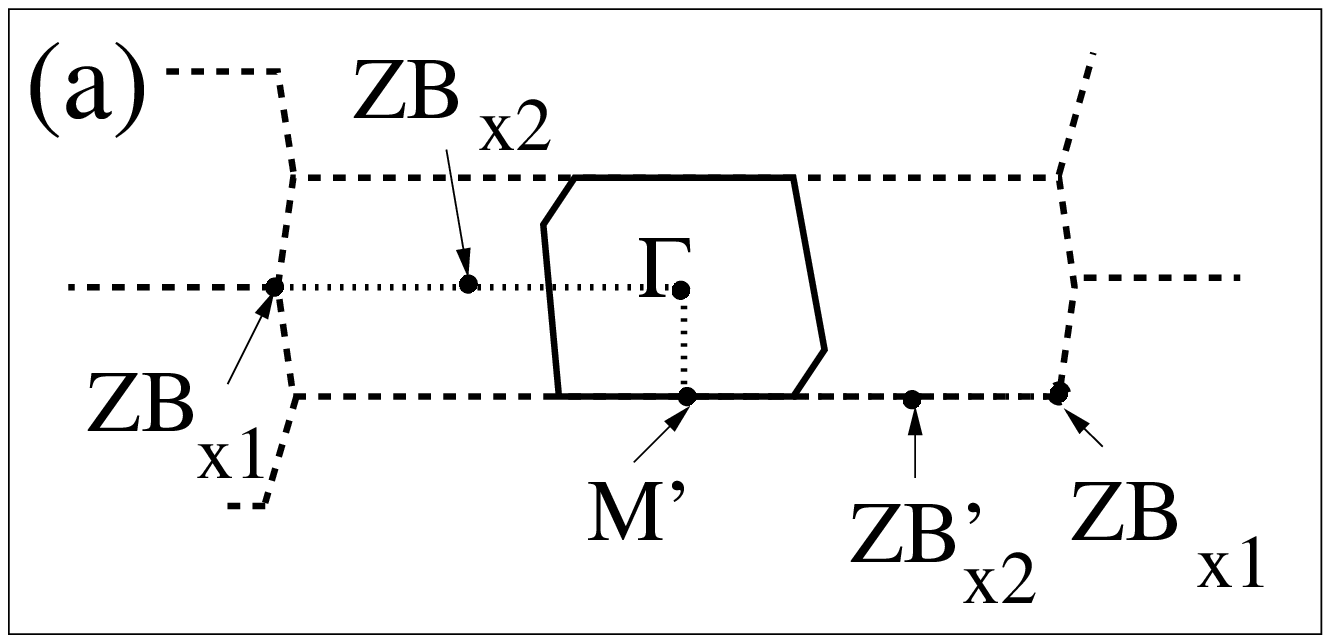}
\end{minipage}
\newline
\begin{minipage}{7.6cm}
\includegraphics[keepaspectratio,width=7.5cm]{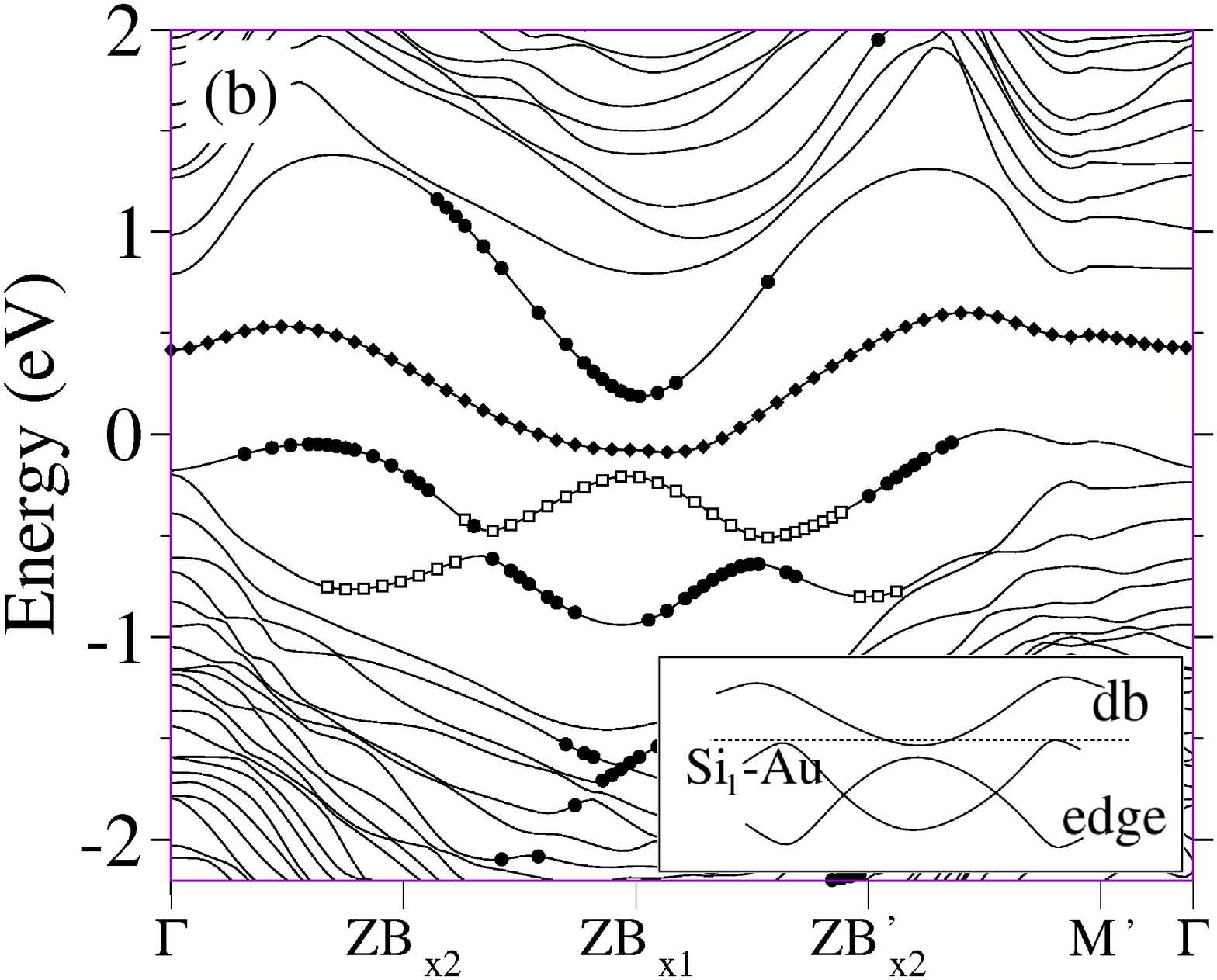}
\end{minipage}
\begin{minipage}{7.6cm}
\includegraphics[keepaspectratio,width=7.5cm]{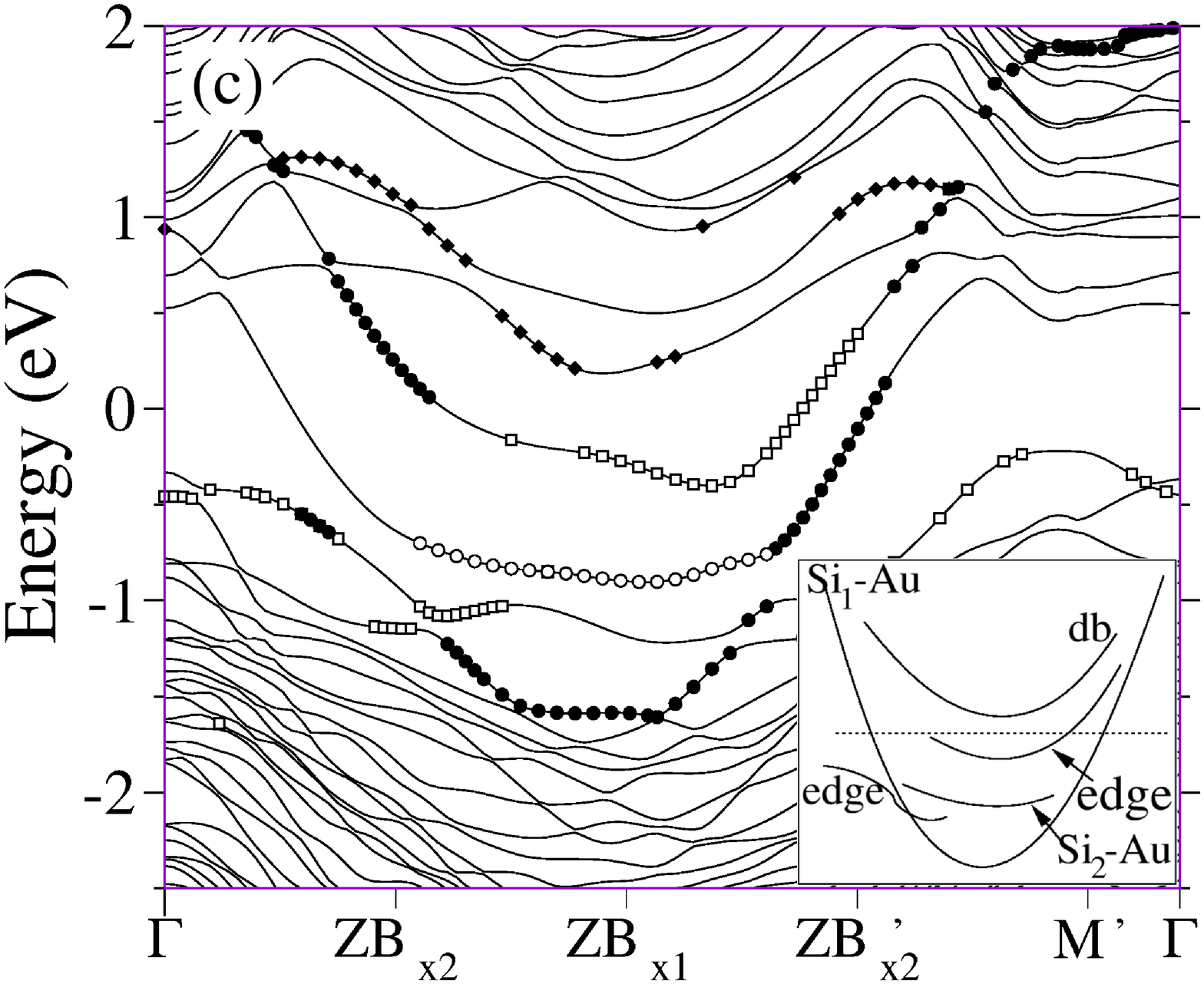}
\end{minipage}
\caption{Panel (a) illustrates the Brillouin zones 
for the ($\times$1) and ($\times$3) periodicities along the step-edge.
The path 
$\Gamma$-ZB$_{\times2}$-ZB$_{\times1}$-ZB$_{\times2}^{\prime}$-M$^{\prime}$ 
is parallel to the steps, while M$^\prime$-$\Gamma$ is perpendicular. 
The band structure of the systems HD$^+$S 
and SS$^+$H are plotted in panels (b) and (c), respectively.
Bands pertaining to the Si$_1$-Au and Si$_2$-Au {\it bonds} 
(solid and open circles, respectively), to the
step edge (squares), and to the dangling bonds 
in the terraces (diamonds) are indicated.
The atomic labels are
consistent with those in figure~\ref{fig:fig2}. 
The insets show a schematic 
representation of the most prominent surface bands. 
\label{fig:fig3}}
\end{figure*}

Figure~\ref{fig:fig3} (b) and (c) show the band structures of
the models I and V, respectively. The different symbols reflect the 
main atomic character of the 
surface bands, as obtained from a Mulliken 
population analysis~\cite{mulliken}.
In figure~\ref{fig:fig3} (b) we can find 
an almost fully occupied band pertaining to the
Si$_1$-Au bonds very close to the Fermi level. 
A dispersive band, coming from the dangling bonds in
the terraces, appears practically empty.
The agreement with the experimental spectra could be improved
if the population of this band would increase~\cite{553}. 
Such electron doping can occur associated with 
the presence of extra silicon atoms 
attached to the step-edge~\cite{himpsel_review}, 
or to the terraces~\cite{doping}.
A surface band associated with the step-edge crosses the Si$_1$-Au 
band
and almost reaches the Fermi level. There is no evidence of this 
band in 
the experiment. The band structure of model II
is almost identical to the band structure of model I.

The band structure of model V (figure~\ref{fig:fig3} (c))
features both a dispersive and a flat
bands associated with the Si-Au bonds.  The step-edge gives rise to
a metallic band with a small filling. The presence of this band implies
that structural distortion 
of the step-edge with an electronic driving force
are possible.
However, as described above, we have failed
to stabilize such distortion in our calculations up to now.

\begin{figure}
\begin{center}
\includegraphics[keepaspectratio,width=8.2cm]{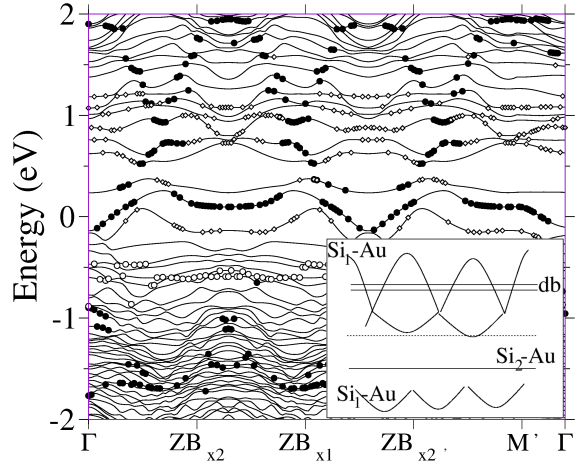}
\end{center}
\caption{ As in figure~\ref{fig:fig3} (b) and (c), 
but for model VI.\label{fig:fig4}}
\end{figure}

The band structure of the model VI is presented in 
figure~\ref{fig:fig4}. The larger unit cell makes the 
comparison with the 
band structures in figure~\ref{fig:fig3} somewhat
complicated due to the backfolding of the bands.
There is a clear gap in the Si$_1$-Au band associated
with the symmetry breaking induced by the silicon dimers 
attached to the step-edges and to some of the atoms in the terraces.  
The changes in the three Au-Au distances in the unit
cell are -0.12, -0.16 and +0.28~\AA, while the corresponding 
Si$_1$-Au-Si$_1$ bond angles
are 103$^\circ$, 100$^\circ$ and 115$^\circ$. Notice that the possible 
mirror symmetry of the structure is also broken in our final 
relaxed configuration. 
This can also be noticed in the slight buckling of
the silicon dimers, visible in figure~\ref{fig:fig1}.

None of the band structures of our models provides a convincing
description of all the available photoemission data for the Si(553)-Au 
surface~\cite{553}.
However, some of the main qualitative features are reasonably
described by the band structure of model V, shown 
in figure~\ref{fig:fig3} (c).  
There are two metallic 
bands with quite similar dispersions.
One of them is close to half-filled and comes 
from Si$_1$-Au bonds (and, therefore, likely to exhibit
an observable spin-orbit splitting). The other one comes from
the step-edge and presents a small fractional filling. 
However, the band originating from the Si$_2$-Au bonds is not seen 
in the experiment and the details of the band structure 
are far from those observed experimentally.

\section{Conclusions}
We presented five possible models for the Si(553)-Au surface constructed
using the silicon honeycomb structure~\cite{honeycomb} and 
the substitution of the gold atoms in the surface
layer.  These models were compared to an earlier 
proposal~\cite{himpsel_review}, which corresponds
to one of our models (model V) with the addition of silicon dimers
at the step edge.
Although the energy differences between different structures
are relatively small, 
one of our models (model I) is clearly the 
most favourable. This
result is stable
against changes in the thickness of the slab and the basis set used
(variations smaller than $\sim$0.5~meV/\AA$^2$). Unexpectedly,
the attachment of extra silicon atoms to the step edge
in model V has little effect on the surface energy. This opens
a route for the modification of the structures
that we plan to investigate in the future for several models.
We have also investigated the possibility to induce periodic modulations
in the structures that might lower the energy and reproduce some
of the 
patterns seen in the STM images~\cite{553}. However, although 
further investigation is also necessary, our structures have proven 
to be quite stable against distortions that would
doubled or tripled the unit cell.

Neither the band structures of our models nor that of the model
proposed by Crain \etal seem to provide a complete description 
of the photoemission spectra. However, the band structure of 
the model V presents some of the main features of the
experimental data.

\section*{Acknowledgments}
This work was 
supported by the Basque Departamento de Educaci\'on,
the UPV/EHU, and
the Spanish MEC, and the ``NANOQUANTA" 
European Network of Excellence.  
SR acknowledges support from the Emil Aaltonen Foundation.
DSP acknowledges support from the Spanish Ministerio 
de Educaci\'on y Ciencia
and CSIC through the ``Ram\'on y Cajal" program.


\section*{References}

\begin{thebibliography}{26}
%
\bibitem{Giamarchi04} Giamarchi T 2004 {\it Quantum
Physics in One Dimension}(Clarendon Press, Oxford)
%
\bibitem{himpsel_review}
Crain J N, McChesney J L, Zheng F , Gallagher M C, Snijders P C,
Bissen M , Gundelach C , Erwin S C and Himpsel F J 2004
{\it Phys. \ Rev. \ B} {\bf 69} 125401
%
\bibitem{luttinger4}
Segovia P, Purdie D, Hengsberger M and Baer Y 1999
{\it Nature} {\bf 402} 504
%
\bibitem{5x2_phase}
Plass R and Marks L D 1997
{\it Surf. Sci.} {\bf 380} 497
%
\bibitem{557_bands}
Losio R , Altmann K N, Lin J L, Petrovykh D Y and Himpsel F J 2001
{\it Phys. \ Rev. \ Lett.} {\bf 86} 4632
%
\bibitem{daniel1}
S\'anchez-Portal D, Gale J D, Garc\'{\i}a A and Martin R M 2002
{\it Phys. \ Rev. \ B} {\bf 65} 81401
%
\bibitem{557_wires}
Ahn J R, Yeom H W, Yoon H S and Lyo L W 2003
{\it Phys.\ Rev.\ Lett.} {\bf 91} 196403
%
\bibitem{sanchez_riikonen}
S\'anchez-Portal D, Riikonen S and Martin R M 2004
{\it Phys. \ Rev. \ Lett.} {\bf 93} 146803
%
\bibitem{adatom_wires}
Yoon H S, Park S J, Lee J E, Whang C N and Lyo I W 2003
{\it Phys. \ Rev. \ Lett.} {\bf 92 } 96801
%
\bibitem{adatoms}
Kirakosian A , Crain J N, Lin J.-L, McChesney J L, Petrovykh D Y,
  Himpsel F J and Bennewitz R 2003
{\it Surf. Sci.} {\bf 532} 928
%
\bibitem{5x2_corr}
Kirakosian A , Bennewitz R , Himpsel F J and Bruch L W 2003
{\it Phys. \ Rev. \ B} {\bf 67} 205412
%
\bibitem{doping}
Erwin S C 2003
{\it Phys. \ Rev. \ Lett.} {\bf 91} 206101
%
\bibitem{memory}
Bennewitz R , Crain J N, Kirakosian A , Lin J L, McChesney J L,
  Petrovykh D Y and Himpsel F J 2002
{\it Nanotechnology} {\bf 13} 499
%
\bibitem{1d_states}
Losio R , Altmann K N and Himpsel F J 2000
{\it Phys. \ Rev. \ Lett.} {\bf 85} 808
%
\bibitem{5x2_bands}
Altmann K N, Crain J N, Kirakosian A , Lin J L, Petrovykh D Y and
  Himpsel F J 2001
{\it Phys. \ Rev. \ B} {\bf 64} 35406
%
\bibitem{arpes_latest}
Matsuda I, Hengsberger M, Baumberger F, Greber T, Yeom
H W and Osterwalder J 2003
{\it Phys. \ Rev. \ B} {\bf 68} 195319
%
\bibitem{553}
Crain J N, Kirakosian A , Altmann K N, Bromberger C , Erwin S C,
  McChesney J L, Lin J L and Himpsel F J 2003
{\it Phys. \ Rev. \ Lett.} {\bf 90} 176805
%
\bibitem{daniel2}
S\'anchez-Portal D  and Martin R M 2003
{\it Surf. Sci.} {\bf 532} 655
%
\bibitem{honeycomb}
Erwin S C.  and Weitering H H 1998
{\it Phys. \ Rev. \ Lett.} {\bf 81} 2296
%
\bibitem{557_xray}
Robinson I K, Bennet P A and Himpsel F J 2002
{\it Phys. \ Rev. \ Lett.} {\bf 88} 96104
%
\bibitem{siesta1}
S\'anchez-Portal D, Ordej\'on P, Artacho E, Soler J M 1997
{\it Int. J. Quantum Chem.} {\bf 65} 453
%
\bibitem{siesta2}
Soler J M, Artacho E, Gale J D, Garc\'{\i}a A, Junquera J, Ordej\'on P,
and S\'anchez-Portal D 2002
{\it J. of Phys.: Condensed Matter} {\bf 14} 2745
%
\bibitem{lda} Perdew J and Zunger A 1981 {\it Phys. Rev. B}
{\bf 23} 5075
%
\bibitem{tm}
Troullier N and Martins J L 1991
{\it Phys. \ Rev. \ B.} {\bf 43} 1993
%
\bibitem{au1}
Hakkinen H , Moseler M  and Landman U 2002
{\it Phys. \ Rev. \ Lett.} {\bf 89} 33401
%
\bibitem{mulliken}
Mulliken R S 1955
{\it J. Chem. Phys.} {\bf 23} 1841
%
\end{thebibliography}
\end{document}